\documentclass[%
 superscriptaddress,
 amsmath,amssymb,
reprint,%
]{revtex4-1}

\usepackage{aeguill}
\usepackage{graphicx}% Include figure files
\usepackage{dcolumn}% Align table columns on decimal point
\usepackage{bm}
\usepackage{url}
\usepackage{verbatim}
\usepackage{lineno}
\usepackage[utf8]{inputenc}
\usepackage{color}
\usepackage{lineno}
\draft % marks overfull lines with a black rule on the right

\begin{document}

%\preprint{}

%\linenumbers

\title{Low energy Ne ion beam induced-modifications of magnetic properties in MnAs thin films} %Title of paper



\author{M. Trassinelli}
\email[]{martino.trassinelli@insp.jussieu.fr}
\affiliation {Institut des NanoSciences de Paris, CNRS, Sorbonne Universités, UPMC Univ Paris 06, 75005 Paris, France}
%\thanks{}
\author{L.~Bernard Carlsson}
\affiliation {Institut des NanoSciences de Paris, CNRS, Sorbonne Universités, UPMC Univ Paris 06, 75005 Paris, France}
\author{S.~Cervera}
\affiliation {Institut des NanoSciences de Paris, CNRS, Sorbonne Universités, UPMC Univ Paris 06, 75005 Paris, France}
\author{M.~Eddrief}
\affiliation {Institut des NanoSciences de Paris, CNRS, Sorbonne Universités, UPMC Univ Paris 06, 75005 Paris, France}
\author{V.H.~Etgens}
\affiliation {Institut des NanoSciences de Paris, CNRS, Sorbonne Universités, UPMC Univ Paris 06, 75005 Paris, France}
\affiliation{Université de Versailles Saint-Quentin en Yvelines, 78035 Versailles, France.}
\affiliation{Institut VEDECOM, 78000 Versailles, France}
\author{V.~Gafton}
\affiliation {Institut des NanoSciences de Paris, CNRS, Sorbonne Universités, UPMC Univ Paris 06, 75005 Paris, France}
\affiliation{Alexandru Ioan Cuza University, Faculty of Physics, 700506 Iasi, Romania}
\author{E.~Lacaze}
\affiliation {Institut des NanoSciences de Paris, CNRS, Sorbonne Universités, UPMC Univ Paris 06, 75005 Paris, France}
\author{E.~Lamour}
\affiliation {Institut des NanoSciences de Paris, CNRS, Sorbonne Universités, UPMC Univ Paris 06, 75005 Paris, France}
\author{A.~Lévy}
\affiliation {Institut des NanoSciences de Paris, CNRS, Sorbonne Universités, UPMC Univ Paris 06, 75005 Paris, France}
\author{S.~Macé}
\affiliation {Institut des NanoSciences de Paris, CNRS, Sorbonne Universités, UPMC Univ Paris 06, 75005 Paris, France}
\author{C.~Prigent}
\affiliation {Institut des NanoSciences de Paris, CNRS, Sorbonne Universités, UPMC Univ Paris 06, 75005 Paris, France}
\author{J.-P.~Rozet}
\affiliation {Institut des NanoSciences de Paris, CNRS, Sorbonne Universités, UPMC Univ Paris 06, 75005 Paris, France}
\author{S.~Steydli}
\affiliation {Institut des NanoSciences de Paris, CNRS, Sorbonne Universités, UPMC Univ Paris 06, 75005 Paris, France}
\author{M.~Marangolo}
\affiliation {Institut des NanoSciences de Paris, CNRS, Sorbonne Universités, UPMC Univ Paris 06, 75005 Paris, France}
\author{D.~Vernhet}
\affiliation {Institut des NanoSciences de Paris, CNRS, Sorbonne Universités, UPMC Univ Paris 06, 75005 Paris, France}

\date{\today}

\begin{abstract}
Investigations of the complex behavior of the magnetization of manganese arsenide thin films due to defects induced by irradiation of slow heavy ions are presented.
In addition to the thermal hysteresis suppression already highlighted in M. Trassinelli \textit{et al.}, Appl. Phys. Lett. \textbf{104}, 081906 (2014), we report here on new local magnetic features recorded by a magnetic force microscope at different temperatures close to the characteristic sample phase transition. 
Complementary measurements of the global magnetization measurements in different conditions (applied magnetic field and temperatures) enable to complete the film characterization.
The obtained results suggest that the ion bombardment produces regions where the local mechanical constraints are significantly different from the average, promoting the local presence of magneto-structural phases far from the equilibrium.
These regions could be responsible for the thermal hysteresis suppression previously reported, irradiation-induced defects acting as seeds in the phase transition.

%New investigations on the ion irradiation effect on manganese arsenide thin films are reported.
%In addition to the thermal hysteresis suppression already highlighted in M. Trassinelli \textit{et al.}, Appl. Phys. Lett. \textbf{104}, 081906 (2014), we report here on the complex behavior of the local magnetization detected by magnetic force microscopy, when irradiation-induced defects are present.
%The microscopic magnetic features are compared to the global magnetization of the film under action of a variable external magnetic field at different temperatures.
%The results suggest that the ion bombardment produces regions where the local mechanical constraints are significantly different promoting the local presence of magneto-structural phases far from the equilibrium.

\end{abstract}

%keyword
%Magnetocaloric effect
%Thermal hysteresis
%Magnetic thin films
%MnAs

\pacs{}% insert suggested PACS numbers in braces on next line

\maketitle %\maketitle must follow title, authors, abstract and \pacs

\section{Introduction}
\label{sec:intro}

Manganese arsenide (MnAs) is one of the most promising materials for the magnetic refrigeration \cite{Gutfleisch2011,Manosa2013,Moya2014}.
It is in fact characterized by a giant magnetocaloric effect close to room temperature that makes it a very attractive material for common refrigeration applications.
The giant magnetocaloric effect of MnAs is related to the large entropy change during the first-order magneto-structural transition close to room temperature ($T_C = 313$~K) between a ferromagnetic $\alpha$-phase with hexagonal structure (NiAs-type) and the paramagnetic $\beta$-phase with orthorhombic structure (MnP-type).  

The possibility of epitaxial growth on standard semiconductors such as GaAs make MnAs thin films interesting also for spintronic research \cite{Daweritz2006} and magneto-elastic applications \cite{Duquesne2012,Marangolo2014}.
The epitaxial strain disturbs the phase transition and leads to the $\alpha - \beta$ phase coexistence over a large range of temperatures (280--320~K).
The refrigeration power associated to the giant magnetocaloric properties does not change but is spread out on the coexistence temperature interval \cite{Mosca2008}.
In this range, an alternating structure of ridges ($\alpha$ phase) and grooves ($\beta$ phase) organized in stripe-shaped domains parallel to MnAs[0001] is created to minimize the elastic energy due to the epitaxy.
The periodicity $\lambda$ of the ridge-groove structure is constant with the temperature and is related to the thickness $t$ of the sample by the relation $\lambda \approx 4.8\ t$ \cite{Kastner2002,Kaganer2002,Breitwieser2009}.

The richness of the MnAs thin films properties results in a complex dependency of the magneto-structural properties on the external temperature, applied magnetic field and constraints.
In the past years, many studies have been dedicated to the characterization of epitaxial MnAs films \cite{Kaganer2000,Kaganer2002,Kastner2002,Daweritz2003,Daweritz2004,Lindner2004,Ploog2004,Adriano2006,Daweritz2006,Garcia2007,Breitwieser2009}.
In particular, magnetic force microscopy (MFM) appeared to be a very useful tool to study the local magnetic properties as a function of different parameters such as: growing conditions \cite{Schippan2000,Daweritz2007}, film thickness \cite{Daweritz2005,Manago2006,Ryu2006}, temperature \cite{Plake2003,Mohanty2003,Mohanty2003a,Coelho2006}, applied magnetic field \cite{Engel-Herbert2005,Engel-Herbert2006} and more recently, applied magnetic field and temperature at the same time \cite{Kim2011}.
In addition to the above investigations based on the out-of-plane magnetic leak field detection, measurement of the in-plane magnetic component imaging has been obtained on cleavage edges \cite{RacheSalles2010}.

Recently it has been demonstrated that slow ions bombardment can interestingly change the thin film properties \cite{Trassinelli2014,Cervera2015}.
New irradiation-induced defects facilitate the nucleation of one phase with respect to the other one in the first-order magneto-structural MnAs transition, with a consequent stable suppression of the thermal hysteresis without any significant perturbation of the other properties. 
Transition temperature, saturation magnetization and structural properties are unchanged  and, more importantly, the giant magnetocaloric properties of the film is maintained.
When present, the thermal hysteresis induces energy losses in the thermal cycles that use giant magnetocaloric material for refrigeration.
Its suppression opens new perspectives for the magnetic refrigeration applications based on MnAs-type material.

The origin of the hysteresis removal being not clear, we present here additional investigations on magnetic properties of the irradiated films under different temperatures and applied field conditions, in particular with the study of the local out-of-plane magnetic field via MFM imaging.
These studies provide new findings that allow to elucidate many aspects of the ion bombardment  effects and also reveal new phenomena.

\section{Film production and irradiation}

All MnAs films considered here have a thickness of 150~nm, are monocrystalline and are issued from the same growth obtained by molecular beam epitaxy (MBE) on GaAs(001) substrate. 
The deposited MnAs is oriented with the $\alpha$-MnAs$[0001]$ and $\beta$-MnAs$[001]$ axis parallel to GaAs$[\bar110]$.
Details on the growth process can be found in Ref.~\citenum{Breitwieser2009}.

The samples are irradiated with a Ne$^{9+}$ ion beam at the SIMPA facility \cite{Gumberidze2010} (French acronym for highly charged ion source of Paris). 
The ion kinetic energy is set at 90~keV (4.5~keV/u) with an incidence angle of 60$^\circ$ with respect to the normal of the sample surface.
These settings were chosen for having the average penetration depth of ions  corresponds to the half-thickness of the MnAs film \cite{Ziegler2008}, without any penetration into the substrate. 
In these conditions, the maximization of the effect arising from the ion irradiation is expected \cite{Zhang2004}.
For the ion energy range considered here, the ion-induced defects are principally produced by the nuclear stopping power with a series of binary collisions between the projectile and the target atoms \cite{Ziegler1985,Averback1998}. 
Target recoil atoms can also produce secondary collisions with the creation of collision cascades that involve many target atoms. 
For one incident atom, about 500 target atom displacements are expected in our conditions.

In the present paper, we focus our investigation on one particular sample that received a fluence $\Phi = 1.5 \times 10^{13}$ ions cm$^{-2}$ and mainly reflect the effects of the ion bombardment in MnAs thin films.
This sample is systematically compared to a non-irradiated one called reference or pristine sample.
More details about the irradiation process can be found in Ref.~\citenum{Trassinelli2014} and  \citenum{ Trassinelli2013}.

\section{Results and discussion}

\subsection{Measurements of the pristine sample}

\begin{figure}
\includegraphics[width=0.65\columnwidth]{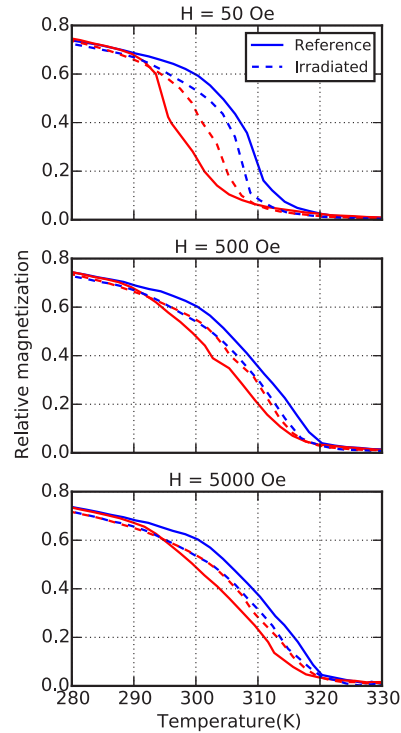}
\caption{Relative magnetization as a function of temperature and external magnetic field for the reference (solid lines) and for the irradiated samples (dashed lines) in proximity of the critical temperature. 
External applied fields are 50~Oe (top), 500~Oe (middle) and 5000~Oe (bottom).
Data obtained by a temperature increase (from colder temperatures) and decrease (from hotter temperatures) are presented in blue and red, respectively.} \label{fig:MT}
\end{figure}

\begin{figure}
\includegraphics[width=\columnwidth]{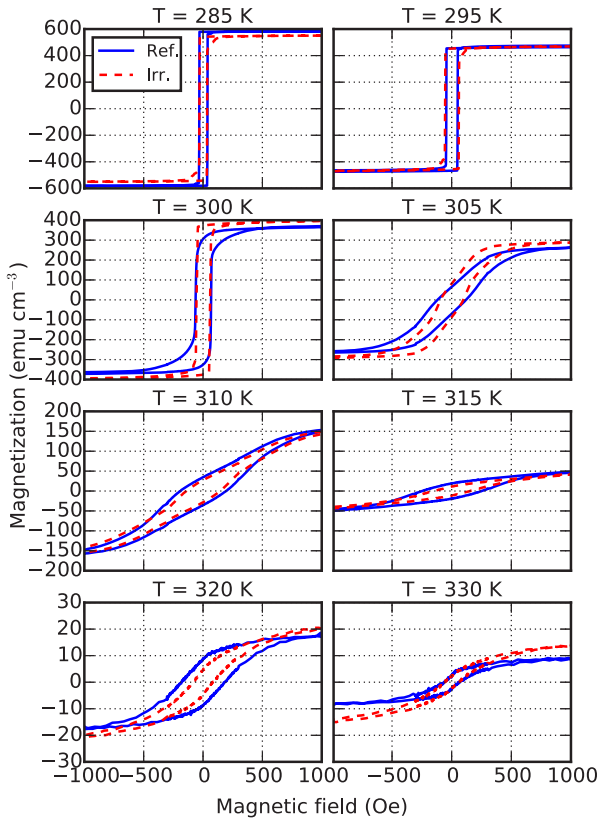}
\caption{Magnetic hysteresis cycles between $H = -1000$ and 1000~Oe for the reference (solid lines) and for the irradiated samples (dashed lines) at different temperatures.} \label{fig:MH}
\end{figure}

The magnetization dependency of the pristine sample with the temperature is mainly driven by the ratio between the $\alpha$ and $\beta$ phases.
For weak applied fields, the reciprocal alignment of the different magnetic domains in the ferromagnetic $\alpha$ phase regions, separated by the paramagnetic $\beta$ phase, plays also an important role.
This effect can be observed in the relative magnetization recorded at different increasing and decreasing values of the temperature shown in Fig.~\ref{fig:MT}.
The sample magnetization has been recorded by a Quantum Design MPMS-XL SQUID magnetometer with a sweep rate of $\pm2$~K/min with 1~T magnetic field applied along the MnAs easy axis $[11\bar20]$.
As visible in Fig.~\ref{fig:MT}, the thermal hysteresis is smaller for intense fields than for weak fields, with $\Delta T^\text{ref} \approx 6.2-6.5$~K for $H=500$ and 5000~Oe and $\Delta T^\text{ref} \approx 11.4$~K for $H=50$~Oe.
This is probably due to the more difficult propagation of the domain walls of ferromagnetic regions separated by paramagnetic regions \cite{Tortarolo2012}.

This interplay between magnetic domains in spatially separated $\alpha$ regions is also responsible of the high coercivity field for the temperatures where the two phases are coexisting (300--320~K).
This is well visible from the sample magnetization measurements along the MnAs easy axis as function of the magnetic field (between $\pm 1$~T) obtained in a VSM magnetometer (Quantum Design PPMS 9T) at different temperatures.
Note that before each magnetic cycle measurement, the history of the sample is erased by magnetic depolarization at $T=350$~K to avoid the production of experimental artifacts like the so-called colossal magnetocaloric effect \cite{Caron2009,Tocado2009}.
As we can observe in the Fig.~\ref{fig:MH},  the domain direction flip takes place at $H \sim 50$~Oe for low temperatures, but increases up to $H \sim 200$~Oe during the phase coexistence. 
This singular behavior is characteristic of MnAs thin films and it has already observed previously \cite{Ney2004,Steren2006,Cervera2015} and successfully modeled \cite{Engel-Herbert2008} in the past.

\begin{figure*}
\includegraphics[width=\textwidth]{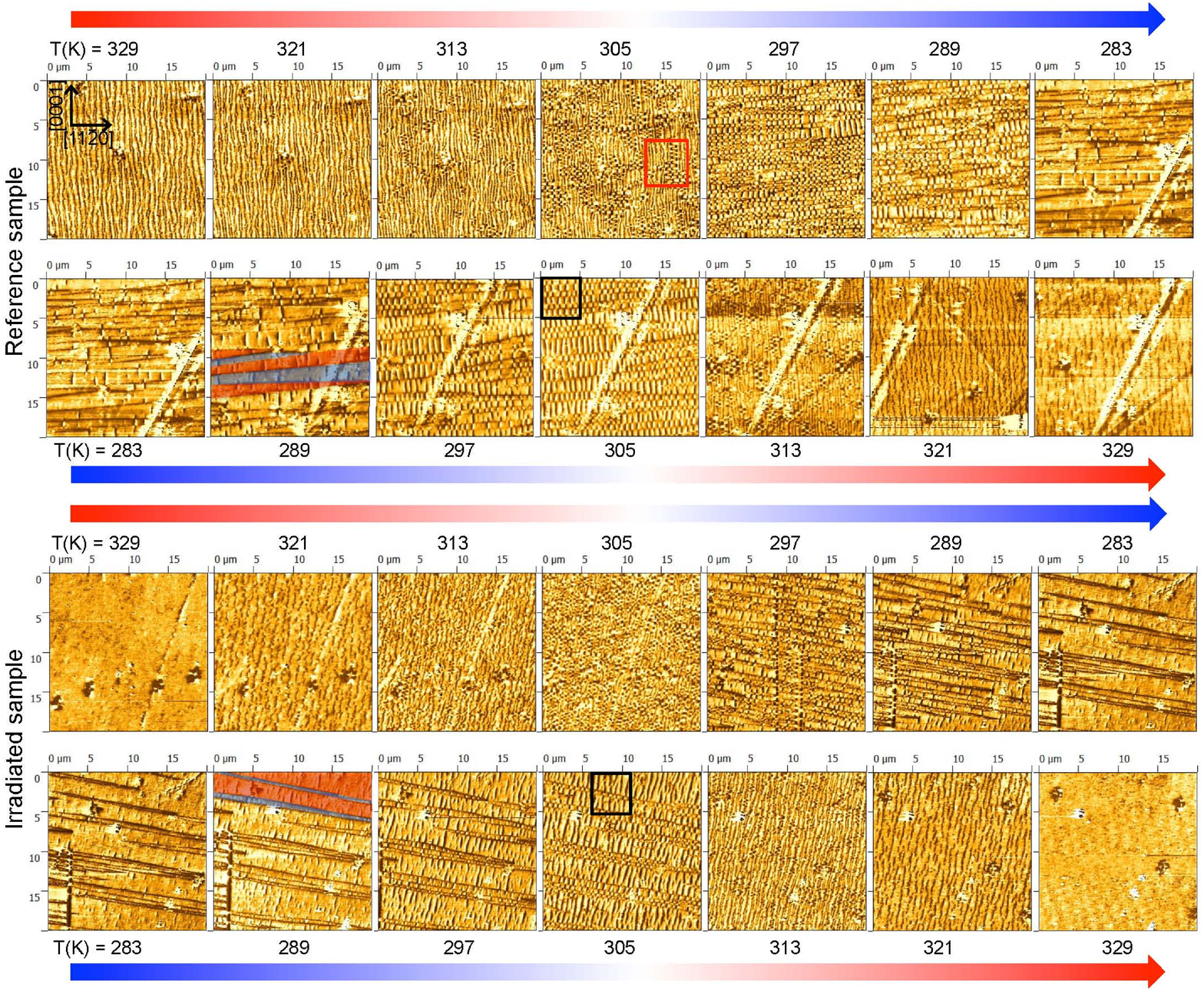}
\caption{MFM images of the reference and irradiated samples at different temperatures.
The colored arrow indicates the temperature progression between the images.
Point-like defects visible at high temperatures are due to growth process defects and they have a correspondence in the topography images.
Similarly, the diagonal line in the lower row of the reference sample images is due to a topological defect. 
In contrast, the diagonal and vertical lines in the irradiated sample images have no correspondence in the topography images and are purely due to magnetic features.
In the image relative to $T=289$~K (increasing temp.), we indicate the macro-domains of opposite magnetization by means of over-imposed blue and red areas.
Squares indicate the zoom regions of Figs.~\ref{fig:types} (in red) and \ref{fig:stripes} (in black).} \label{fig:MFM}
\end{figure*}

\begin{figure}
\includegraphics[width=0.8\columnwidth]{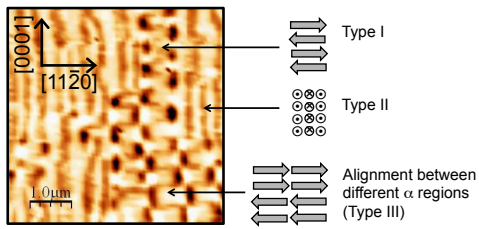}
\caption{Different types of magnetic orientation of the $\alpha$-phase regions in a reference sample MFM image at $T=305$~K.
The image correspond to the zoom of the red square in Fig.~\ref{fig:MFM}.} \label{fig:types}
\end{figure}

The features of the MFM images obtained at different temperatures and no external field (Fig.~\ref{fig:MFM}, top) reflect the complicate interplay between magnetic domains. 
The direct observation of the $\alpha$- and $\beta$-phase regions layout is obtained with a MFM (Bruker Multimode AFM microscope equipped with a magnetic tip coated with Co/Cr, model MESP, with a lift height of 20~nm) equipped with a Peltier regulator to adjust the temperature. 
The measurement is performed under normal atmospheric pressure and it was limited to temperatures higher than the dew point ($T \sim 280$~K) to avoid water condensation on the sample. 
The different images are acquired starting at decreasing and subsequently increasing sample temperatures, from the initial value of $T=329$~K.
During the temperature change, spatial drifts of the sample did not allow for the measurement of very high quality images. In particular it did not allow for a perfect correspondence between the different spatial regions in of a given area.

Because of the absence of applied external field, MFM images correspond to the visualization of the out-of-plane magnetic field component from $\alpha$-phase regions separated by $\beta$-phase regions.
Different temperatures correspond to different typical sizes of the $\alpha$-phase regions that favor different domain types that are presented in Fig.~\ref{fig:types} (see also \cite{Plake2003,Engel-Herbert2005,Coelho2006,Daweritz2007,Kim2011}).
When the ferromagnetic regions are sufficiently extended, the magnetic domains are normally aligned along the easy magnetization axis MnAs$[1 1 \bar 2 0]$ with the formation of type-I magnetic domains, with meander-like contrast.
We can also have entire portions of $\alpha$-phase regions in a single domain state (type-III domains) aligned to other ferromagnetic regions orientation across different $\beta$-phase regions, forming what we call macro-domains.
For very small $\alpha$-phase regions, type-II domains,  with magnetic moment perpendicular to the surface, are present.

As visible in Fig.~\ref{fig:MFM}, when coexistence of $\alpha$ and $\beta$ phases exists, those regions are elongated in fact along the MnAs[0001] direction, which is perpendicular to the easy magnetization axis (MnAs$[1 1 \bar 2 0]$).
Point-like features, well visible at high temperatures, are due to growth process defects and they have a correspondence in the topography images.
Similarly, the diagonal line in the lower row of the reference sample images is due to a topological defect. 

Because of the reduced extension of $\alpha$-phase regions, type-II domains are dominant at high temperatures.
When more extended regions of $\alpha$-phase are present ($T=283-297$~K),  type-I domains start to appear and form macro-domains.
As shown in Fig.~\ref{fig:MFM}, large domains are recognizable as elongated zones aligned width the easy magnetization axis MnAs$[11 \bar 20]$ exhibiting black-white stripes alternation typical of MnAs MFM images during the phase coexistence \cite{Daweritz2006}.
To facilitate the readability of the MFM images, we indicate in the image relative to the reference sample at $T=289$~K in Fig.~\ref{fig:MFM} the extension of the macro-domains of opposite magnetization by means of over-imposed blue and red areas.
The limits between different macro-domains are easily recognizable by the flip of the MFM phase contrast of the stripes. 
At lower temperatures ($T=283$~K), $\beta$-phase regions disappear completely on large portions of the films leaving MFM image regions without any contrast.

\begin{figure}
\includegraphics[width=0.8\columnwidth]{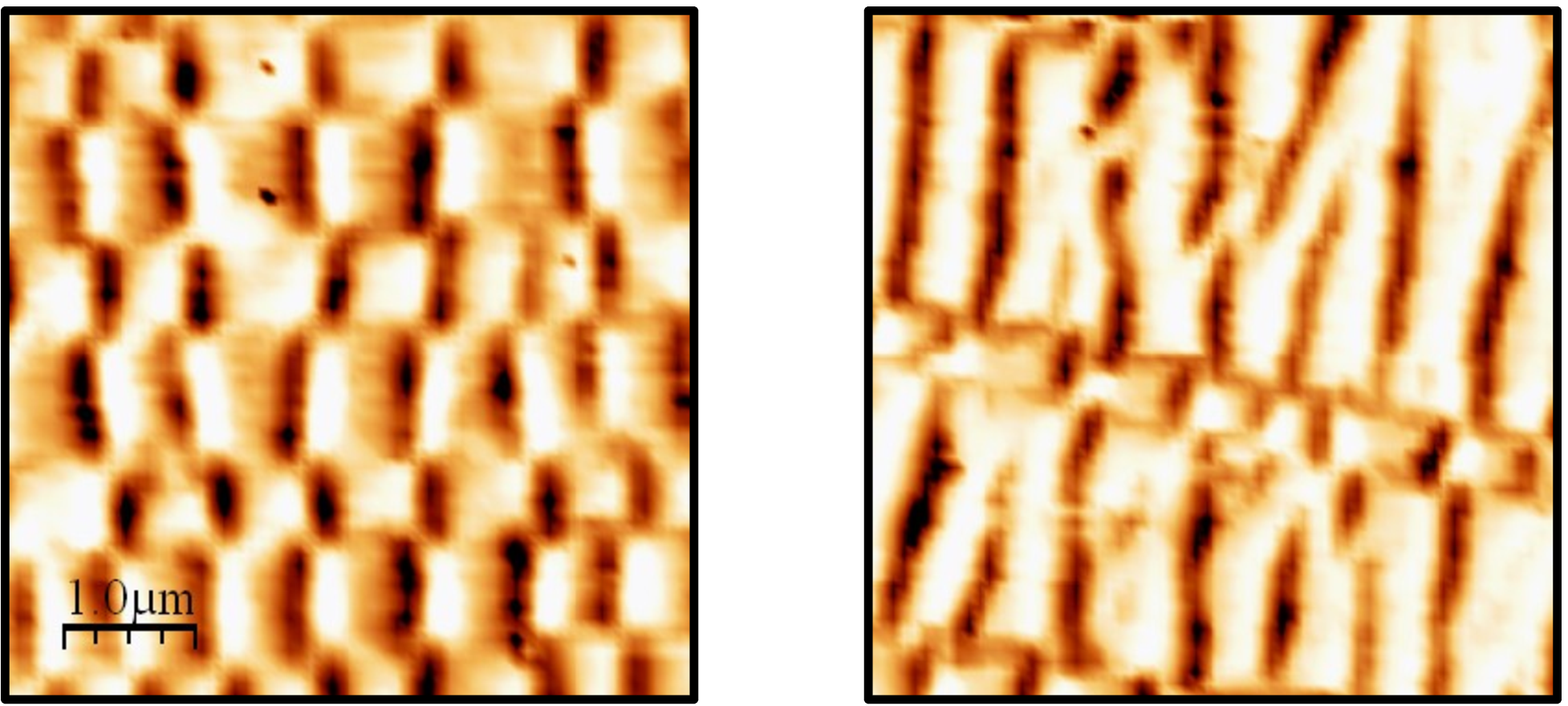}
\caption{Zoom of the topography images at $T=305$~K of the pristine (left) and irradiated (right) samples.
These images correspond to the zoom of the black squares in Fig.~\ref{fig:MFM}.} \label{fig:stripes}
\end{figure}

When the temperature increases, regularly spaced stripes reappear at $T=297-305$~K (Figs.~\ref{fig:MFM} and \ref{fig:stripes}). 
At $T\sim 313$~K, type-I and II domains reappear, and only type-II domains are present at higher temperatures. 

\subsection{Effect of the ion irradiation}
Irradiation of the thin film with slow ions produce new defects in the MnAs crystal.
As reported in Refs.~\citenum{Trassinelli2014,Cervera2015}, these new defects act as nucleation seeds of one phase with respect to the other during the magneto-structural transition leading to the elimination of the thermal hysteresis.

Investigations with X-ray diffraction for possible structural changes show that the $\alpha$ and $\beta$ phases structure of irradiated samples, up to a fluence of $10^{15}$ ions cm$^{-2}$, are unchanged as presented in Refs. \citenum{Trassinelli2014,Cervera2015}.
The film remains monocrystalline and only a small increase of the average lattice spacing is noticed (less than 0.3\% for a fluence of $1.6 \times 10^{15}$~ions cm$^{-2}$).

In addition to the thermal hysteresis suppression previously reported, irradiation-induced modifications are additionally characterized by magnetometry measurements for different applied magnetic field values ($M(H)$) and temperatures ($M(T)$), and by MFM imaging.
These measurements provide new insights on the nature of the bombardment damages.
As we can see in the magnetization measurements as a function of the temperature in Fig.~\ref{fig:MT}, for $H=500, 5000$~Oe, the major difference between the pristine and irradiated samples is the suppression of the thermal hysteresis without variation of the critical temperature $T_c$ of the transition.
This is not true when a magnetic field of 50~Oe is applied.
There, the thermal hysteresis is still present and is $\Delta T^\text{irr} \approx 4.8$~K (Fig.~\ref{fig:MT}).
Nevertheless, we note that the difference between $\Delta T^\text{ref} - \Delta T^\text{irr}$ has a constant value of  about $6.2-6.6$~K for any applied field.
This suggests that for $H=50$~Oe, the non-zero value of $\Delta T^\text{irr}$ is not due to additional pinning caused by irradiation-induced defects but most likely to an intrinsic property of MnAs/GaAs film.

More information can be extracted from the magnetization measurements as a function of the applied magnetic field  at different temperatures (Fig.~\ref{fig:MH}).
No major differences in the shape of $M(H)$ between reference and irradiated samples are visible except at the temperatures $T=300$ and 305~K where the inflection point position or the slope changes.
In these cases, referring to Fig.~\ref{fig:MT},  one can deduce that the $\alpha/\beta$ phase ratio plays a role due to the thermal hysteresis.

As previously reported \cite{Trassinelli2014,Cervera2015}, the ion irradiation does not affect the coercivity field values.
Changes of the saturation magnetization are however visible. 
At $T=285$~K the absolute value of the magnetization of the irradiated sample is slightly lower than in the reference one.
This is also visible in the $M(T)$ measurement in Fig.\ref{fig:MT}.
But no noticeable difference of the magnetic saturation is visible at lower temperature  ($T=100$~K). 
A possible explanation of this behavior is the presence of residual $\beta$-phase regions when only the $\alpha$ phase should be present. 
This hypothesis is confirmed by the MFM images that are discussed in the next paragraph.
Finally it is worth mentioning that thin film magnetization reduction under action of ion irradiation is also commonly observed in samples that exhibit a second-order magnetic transition because of irradiation-induced crystal structure rearrangement and change of the dipole-dipole interaction \cite{Weller2000,Kaminsky2001,Zhang2003,Fassbender2006,Fujita2010,Cook2011,Tohki2012,Oshima2013}.
However, compared to the results presented here, these past studies involved much higher ion fluences ($> 10^{15}$ ions cm$^{-2}$) that, differently than in our case, can substantially change the sample structure and composition.

Additional insights can be extracted from the magnetic force microscope images and topography of the sample.
The topography imaging does not show any differences between the pristine and irradiated sample.
This is not surprising because the relative low-value of the potential energy carried by the ions does not allow the formation of new surface defects \cite{Aumayr2011}.
On the other hand, compared to the reference sample, the MFM images (Fig.~\ref{fig:MFM}) of the irradiated sample present several differences:
\begin{enumerate}
\item there is a qualitatively different dependency on the temperature, \label{it:temp}
\item they are characterized by an irregular pattern of the $\alpha-\beta$ phase region disposition, \label{it:irr}
\item at low temperature, persisting  $\beta$-phase regions are observed. \label{it:fossile}
\end{enumerate}

Point \ref{it:temp} is due to the reduction or suppression of the thermal hysteresis that anticipates the appearance of one phase with respect to the other one in the irradiated sample compared to the pristine one.
As discussed above, this reduction is caused by the easier nucleation of the phases due the irradiation-induced defects.
These defects disturb also the regular $\alpha-\beta$ phase disposal, point \ref{it:irr}, as clearly visible in Fig.~\ref{fig:stripes}.
This phenomenon is similar to the effect of defects produced during the film growth that are observed to induce Y-shape stripe bifurcation in well defined positions (corresponding to the defect positions) \cite{Daweritz2003,Breitwieser2009}.

\begin{figure}
\includegraphics[width=0.8\columnwidth]{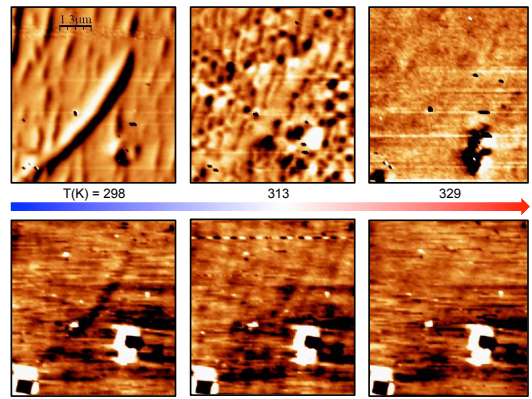}
\caption{MFM (upper row) and AFM (bottom row) images at different increasing temperatures of the irradiated sample after a cooling in presence of a persisting magnetic field ($H \sim 5000$~Oe).} \label{fig:beta}
\end{figure}

Point~\ref{it:fossile} is particularly interesting.
At low temperature, we can see in Fig.~\ref{fig:MFM} that the macro-domains are disturbed by linear regions where the $\beta$ phase is persisting. 
These regions are recognizable by a local contrast of the MFM phase that develop along lines and by a depression of a few nm in the topography lines (not shown) due to the smaller cell volume of the $\beta$ phase \cite{Daweritz2006,Breitwieser2009}.
These regions are particularly visible in the image relative to $T=297$~K approaching from higher temperatures in Fig.~\ref{fig:MFM}.
For lower $T$ values, only one vertical $\beta$-phase region is persisting, down to the lowest reached temperature of 283~K.
This effect is even more evident when the sample is cooled from about 350~K with an external field of about 5000 Oe before imaging it with the MFM.
As we can see in Fig.~\ref{fig:beta}, where a $5 \times 5~\mu$m$^2$ of the irradiated sample at the temperature of 298~K prepared in this way is shown, a $\beta$-phase diagonal region is visible in the MFM image together with a depression (of about 5~nm) in the topography image.
When increasing temperature, this $\beta$-phase region is less and less contrasted with respect to the surrounding area in both MFM and topography images due to the global transition to the $\beta$ phase.

The presence of the $\beta$ phase at this low temperature may be due to local constraints induced by the ion irradiation.
Past experimental and theoretical studies on ion--matter interaction show that, at this ion energy regime, collision cascades can lead to the formations of spatially localized regions rich in interstitial atoms or vacancies \cite{Averback1998,Nordlund1999,Kim2012,Lu2016}.
Interstitial-rich regions could cause an increase of the local high internal pressure favoring the presence of the $\beta$ phase that is characterized by a volume 2\% larger than the $\alpha$ phase.
In connection to the persistence of small $\beta$-phase regions, the sample magnetization is expected to be reduced with respect to the reference sample, which is the case as visible in the magnetometry measurements in Fig.~\ref{fig:MT}.
At very low temperature ($T=100$~K) not monitored with the MFM, irradiated samples have practically the same saturation magnetization than pristine samples \cite{Trassinelli2014} suggesting that these ``frozen'' $\beta$-phase regions are thus suppressed.

Similarly, the presence of surviving $\alpha$-phase regions at high temperatures seems plausible, but is difficult to observe due to the complex magnetic images caused by the transition between type-I and type-II magnetic domains.
In correlation with vacancies-rich regions, a local low internal pressure is expected that can favor the presence of $\alpha$ phase regions.
The high magnetization of irradiated samples at $T=330$~K in the $M(H)$ curve (Fig.~\ref{fig:MH}) could be an indication of the persistence of $\alpha$-phase regions at high temperatures.
This persistence could be the reason for the presence of large macro-domains with the same orientation contrary to what happen in the pristine sample (Fig.~\ref{fig:MFM},  $T=283-297$).
Remaining $\alpha$-phase regions with a defined magnetic orientation could in fact seed the nucleation of domains with a specific direction (before the series of observation on the MFM the samples are initially cooled in presence of a persisting magnetic field of about 5000~Oe).

Recent studies on nanoindentation in NiMnGa Heusler alloy films \cite{Niemann2016} show that local constrains can favor the nucleation of one phase into the other phase in the characteristic first-order phase transition.
Similarly, here local regions of $\alpha$ and $\beta$ phase, which are persisting at a temperature respectively higher and lower than the transition temperature, could be at the origin of the thermal hysteresis suppression making easier the transition of one phase into the other phase.

\section{Conclusions}
We report new investigations on the effect of ion bombardment on manganese arsenide thin films.
These studies confirm that the thermal hysteresis is suppressed when an intense magnetic fields is applied due to local irradiation-induced defects.
This is not the case for weak external field. At 50~Oe the surviving thermal hysteresis seems related to the intrinsic properties of MnAs/GaAs films and in particular to the high coercivity field value during the $\alpha-\beta$ phase coexistence.
The local effects of the new defects are well visible on the MFM images where the self-organization of MnAs $\alpha-\beta$ phases is disturbed and where $\beta$-phase regions persist in specific spots at temperatures well below the phase transition temperature.
This is in agreement with the observed magnetization of the irradiated sample which is lower than the pristine magnetization for $T \lesssim T_c$ while at lower temperature, i.e. $T=100$~K, the magnetization is the same for the two samples.
Similarly, there are indications that small $\alpha$-phase regions could resist to temperatures higher than the phase transition temperature.
Nevertheless the nature of the irradiation-induced defects is still not completely revealed and it will require future investigation with different projectile masses and collision conditions. 

\section*{Acknowledgments}
This work was partially supported by French state funds managed by the ANR within the Investissements d'Avenir program under reference ANR-11-IDEX-0004-02, and more specifically within the framework of the Cluster of Excellence MATISSE led by Sorbonne Universités. 
V. Gafton acknowledges support from the strategic grant  POSDRU/159/1.5/S/137750, Project, ``Doctoral and Post-doctoral programs support for increased competitiveness in Exact Sciences research'' co financed by the European Social Found within the Sectoral Operational Program Human Resources Development 2007-2013

\bibliography{MFM_and_co2}

\end{document}